\documentstyle[12pt,aaspp4]{article}
\lefthead{Hurley-Keller, Mateo, \& Grebel}
\righthead{Second Parameter Problem in Sculptor}
\def\lsim{\mathrel{\lower .85ex\hbox{\rlap{$\sim$}\raise
.95ex\hbox{$<$} }}}
\def\gsim{\mathrel{\lower .80ex\hbox{\rlap{$\sim$}\raise
.90ex\hbox{$>$} }}}

\begin{document}

\title{A New Culprit in the Second Parameter Problem in the Sculptor Dwarf
Spheroidal Galaxy?}

\author{Denise Hurley-Keller\altaffilmark{1} \& Mario Mateo\altaffilmark{1}}
\affil{Astronomy Department, University of Michigan, Ann Arbor, MI 48109; denise@astro.lsa.umich.edu, mateo@astro.lsa.umich.edu}
\altaffiltext{1}{Visiting Astronomer, Cerro Tololo Inter-American Observatory, National Optical Astronomy Observatories, operated by AURA, Inc., under cooperative agreement with the National Science Foundation.}
\author{Eva Grebel\altaffilmark{2}}
\affil{University of Washington, Department of Astronomy, Box 351580, Seattle, WA 98195; grebel@betula.astro.washington.edu}
\altaffiltext{2}{Hubble Fellow.}
\singlespace
\begin{abstract}
\singlespace

Color-magnitude diagrams from deep, wide-field CCD photometry of the
Sculptor dwarf spheroidal galaxy reveal that the red horizonal branch
(RHB) stars are strongly concentrated towards the center of the galaxy
relative to the dominant old population in Sculptor, confirming an
earlier claim of such a gradient (Da Costa et al. 1996).  Since we
find no radial gradients of the age or metallicity distribution within
Sculptor, neither age nor metallicity can individually account for the
internal `second parameter' problem observed within the galaxy.  We
have also identified an unusual `spur' of stars that extends from the
main sequence turnoff region and located between the canonical blue
straggler region and the subgiant branch.  These stars are also more
centrally concentrated than the oldest stars in Sculptor, although not
as extremely as the RHB stars.  Unable to convincingly interpret the
spur either as an unusual young population or as a foreground
population of stars, we conclude that binary stars offer the most
reasonable explanation for the origin of the spur that is also
consistent with other features in the CMD of Sculptor. We infer that
30-60\%\ of all apparently single stars in the inner region of
Sculptor may be binaries.  We speculate that the possible radial
gradient in the binary-star population may be related to the variation
of the HB morphology in Sculptor.

\end{abstract}


\newpage
\section{Introduction}
\baselineskip=25pt 
\singlespace
		
The dwarf spheroidal (dSph) companions of the Milky Way have proven to
be useful laboratories to study the remarkable range of star-formation
histories (SFH) that can occur on the smallest galaxian scales
(e.g. Carina, Fornax, Leo~I; see Mateo 1998 and references therein;
also Grebel 1997).  Ironically, the first dSph galaxy to be discovered
-- Sculptor (Scl; Shapley 1938) -- has long been believed to have
experienced the relatively sedate star-formation history (SFH) that
was initially attributed to all dSph galaxies (Da~Costa 1984; Kaluzny
et al. 1995).  The recent discovery of neutral gas associated with Scl
(Carignan et al. 1998) suggests a more complex history.  Past
photometric studies of Scl have either been too shallow or covered too
little area to study the spatial variations of {\it all} stellar
populations in the galaxy.  We therefore set out to obtain deep,
wide-field CCD photometry to derive a complete SFH of Scl
(c.f. Stetson et al. 1998), particularly in regions near the
newly-discovered gas.  A second paper will present detailed results of
a comprehensive analysis of the SFH of Scl (c.f. Hurley-Keller et
al. 1998).

This {\it Letter} is motivated by two unexpected findings.  First, we
observe that the red horizontal-branch (RHB) stars in Scl are
significantly more centrally concentrated than any other stellar
component of the galaxy, confirming earlier suggestions of such a
gradient (Da~Costa et al. 1996; Majewski et al. 1999).  Second, we
have identified an unusual sequence of stars that extends from the
main-sequence turnoff (MSTO) between the region containing blue stragglers
(BS)/intermediate-age stars and the subgiant branch stars.  We
describe these features and offer a preliminary discussion of their
possible origins and implications.

\section{Observations}

The photometry was obtained with the Big Throughput Camera (BTC;
Wittman et al. 1998; see {\tt www.astro.lsa.umich.edu/btc/tech.html}
for details) at the CTIO 4m telescope on 1998 September 24.
The camera contains four CCDs that each cover $14.7' \times 14.7'$
covering a total field of about 0.25 deg$^2$ with a pixel size of
0.43$\arcsec$.  For the principal observations discussed here, CCD 4
(located at the SW corner of the BTC) was centered on Scl and
covered the entire inner region.  The remaining three CCDs were
centered 20-28 arcmin from the center of the galaxy, well outside the
core radius of 5.8 arcmin (Irwin \& Hatzidimitriou 1995).  We refer to
these observations as `pointing 1'.  We also obtained a few
frames with Scl centered on CCD 2; we refer to these as `pointing 2'.
All data were obtained using the standard CTIO Johnson $B$ and 
Kron-Cousins $R$ filters.

We applied standard processing and reduction techniques (with DoPHOT;
Schecter et al. 1993) to the images.  The present results are based on
photometry measured from deep images produced from the sum of all
exposures in each filter (a total of 7200 sec in each filter).  We
have 10 stars in common with Da~Costa (1984) results; our photometry
is systematically consistent with this earlier study.  For unknown
reasons, the photometry from CCD 2 is significantly poorer than for
the other CCDs. Apart from one test described below, we have chosen
not to use the photometry from that chip in this study.

\section{The Color-Magnitude Diagrams}

Figure~1 shows the CMDs of the central $\sim15' \times 15'$ region of
Scl (from CCD 4), and of the outer region (combined results from CCDs
1 and 3).  Nearly 70,000 stars are present in the `inner' CMD and
about 20,000 in the `outer' CMD.  Both CMDs exhibit a blue horizontal
branch (BHB), a red giant branch (RGB), a well-defined subgiant branch
(SGB), an old ($\sim15$ Gyr) MSTO, and a significant number of
relatively blue, luminous stars which could either be canonical blue
stragglers (see Piotto et al. 1999 for examples) or intermediate-age
stars in Scl.

The CMDs show that the well-populated red horizontal branch (RHB)
observed in the central region has essentially vanished in the outer
region.  Da~Costa et al. (1996) have observed similar HB gradients in
other dSph systems, and had suggested based on the Gunn system CCD
photometry of Light (1988) that the HB population in Scl exhibits such
a gradient.  This effect has also been reported by Majewski et
al. (1999).  Our new data unambiguously confirm the existence and
sense of the gradient (see Figure~1).  The central CMD also clearly
contains a feature unlike any seen in deep CMDs of other dSph
galaxies: a well-populated `spur' of stars extending $\sim 0.7$ mag
above the old MSTO.

We can quantify the spatial distributions of the Scl populations using
star counts.  Figure~2a illustrates the locations of five regions
sampling distinct populations within the CMD, as well as a sixth
region in which false stars were added (see section 4).  The
boundaries of the BS, Spur, and SGB boxes ensure that the relative
completeness in each one are similar.  Table~1 lists the star counts
in the CMD boxes as well as relevant ratios of the star counts.
Because the CMDs of the outer CCDs do not differ significantly (CCD 2
excepted), the star counts for chips 1 and 3 were combined to improve
the signal-to-noise of the outer-region counts.  We assume Poisson
uncertainties throughout this paper, and we consider a result to be
`significant' if the compared values differ by more than three times
their combined uncertainties (i.e. $\Delta \ge 3$). We also employ the
T2 statistic used by Da~Costa et al. (1996) to compare count ratios.
Table~1 confirms the visual impression of Figure~1; relative to the
old subgiant stars, there are five times as many RHB stars in the
center as in the outer regions (RHB/SGB).  The spur stars are the only
other population to reveal any evidence of an over-concentration in
the center of Scl.

The gradient in the HB morphology is not due to a simple age or
metallicity gradient between the inner and outer regions.  The
magnitude and color of the old MSTO in both CMDs are the same to
within 0.05 magnitudes, allowing at most a 1-2 Gyr variation in the
mean age of the older turnoff component for a constant metallicity.
While the color spread of the SGB reveals a clear metallicity
dispersion (confirming the finding of Majewski et al. 1999), we find
no evidence that the {\it mean} color of SGB stars differs between
the inner and outer regions by more than 0.02 mag.  Nor does the shape
of the distribution vary; the ratios of blue to red SGB stars between
the inner and outer regions show no statistically significant gradient
from the inner to outer fields.  We conclude that neither an age nor a
metallicity gradient can individually account for the observed radial
gradient of the HB morphology of Scl.

\section{The Nature of the Spur}

We have explored a number of interpretations to account for the spur
stars, which have never been seen in any other dSph system and which
may be related to the RHB gradient.

\noindent {\it 1. Are the spur stars photometric blends?}  The median
FWHM of spur stars in the inner field is $\sim 1.2$ arcsec, identical
to that of SGB and BS stars of similar brightness in that field; their
appearance is also similar. They are distributed in the same way as
all other stars and so are not associated with some particular
structure or defect in the CCD.  We also reduced the shallower,
noisier data for pointing 2.  The resulting CMD of the inner field has
the same features as that derived from pointing 1.  The same stars
located in the spur in pointing 1 are still spur stars in pointing 2.

We added to the images 1500 false stars (100 stars at a time) drawn
from an isochrone corresponding to the oldest population in Scl and a
Salpeter-like IMF.  The region labeled "Old" in Figure 2a samples a
portion of this isochrone. We then compared the number of false Spur
stars to the number of false Old stars.  For every 100 stars from the
false star sample detected in the Old region, 7 objects were detected
in the Spur region; for the data, 19 Spur stars were detected for
every 100 Old stars.  At that rate, photometric blends could only
account for a minority of the real spur stars ($\sim$30-40\% = 7/19).
Unlike the genuine spur stars, visual inspection of the false spur
star images clearly shows them all to be photometric blends.  We
conclude that most of the spur stars of the inner field are not the
result of image blending or other instrumental effects.

Similar tests in the outer region produced only 3 spur stars for every
100 Old stars.  If we correct the inner and outer Spur/SGB proportions
for blends, we still dectect a radial gradient of the relative number
of spur stars based on both statistical criteria in Table~1.  While
the gradient in spur star counts (only a factor of $\sim2$) is
substantially less than that of the RHB stars, this is nonetheless
circumstantial evidence that the spur may play some role, albeit
uncertain, in producing the strong HB gradient in Scl.

\noindent {\it 2. Is the spur a sequence of young stars in Scl?}
Figure~2b compares our data with a variety of Padua isochrones
(Bertelli et al. 1994) plotted to `fit' the oldest main sequence stars
and the stars in the BS region.  We assumed a metallicity of
$z=0.0004$, a solar abundance distribution, and a distance modulus of
19.5, values consistent with previous studies of Scl (Da~Costa 1984;
Grebel et al. 1994; Kaluzny et al. 1995; Majewski et al. 1999).  We
neglect the foreground extinction towards Scl ($A_V \lsim 0.05 \pm
0.05$ mag; Mateo 1998).

The oldest stars in Scl can be fit with models corresponding to ages
of about 14-15 Gyr, while the BS region can be adequately described as
a sequence of intermediate-age stars older than about 4-5 Gyr.  We
then tried to fit an intermediate-age model to the spur. For the same
metallicity/distance parameters, we found that the $R$-band luminosity
of the `turnoff' of the spur corresponds to an age of about 6-8 Gyr.
The corresponding isochrone represents the color of the spur very
poorly (Fig. 2b).  If we instead match the spur's `turnoff' color, we
require a more metal-rich, younger isochrone.  This model (dashed
line) predicts a sparse RGB to the red of the principal RGB, similar
to a feature seen in the CMD of the Sagittarius dSph galaxy
(Sarajedini \& Layden 1995), which is certainly absent in Scl.  We
conclude that the spur is not the turnoff region of a normal
intermediate-age stellar population.

\noindent {\it 3. Is the spur associated with a foreground object?}  A
15 Gyr isochrone with z=0.0004 but an assumed distance modulus of 18.8
fits the spur reasonably well (Figure~2c).  Unfortunately, this good
fit is achieved at the cost of a highly contrived scenario.  We would
require that the putative foreground object has an old stellar
population identical to that of Scl, but is located $\sim 20$ kpc in
front of the galaxy directly along the line of sight between Earth and
the center of Scl.  If the foreground component were aligned instead
along the line connecting Scl and the Galactic center (Piatek \& Pryor
1995; Kroupa 1997), we would expect it to be offset by up to 2 degrees
from the center of Scl as seen from the Sun.  There is also no hint of
a corresponding red or blue HB component located $\sim 0.7$ mag above
the observed HB stars in Scl that could be associated with the old
component of the foreground system.

\noindent {\it 4. Is the spur a photometric binary sequence?}  We used
the population synthesis code from Padua (Bertelli et al., private
communication) to carry out preliminary simulations of a binary star
population in Scl.  Input parameters include the binary fraction ($f$,
the fraction of apparently single stars that are truly binary systems)
and the range of binary mass ratios ($q$, where $0.0 < q \leq 1.0$).
The stars which become binaries and their masses are chosen randomly
from uniform distributions of these parameters.  The models do not
reproduce effects from binary interactions or mass transfer.

Our simulations adopted $f$ in the range 0.30 to 0.95 for binary
populations with $q \ge 0.7$ to $q \ge 0.95$.  The best fits to the
data require $f \sim 30$-60\%\ and a distribution of mass ratios with
most binaries having $q \gsim 0.7$.  The quality of the fits was
determined by comparing the color distribution and luminosity function
of the data and the models in the spur region, and across the subgiant
branch (e.g. Hurley-Keller et al. 1998).  Models with $f \gsim 0.6$,
for example, begin to predict too many spur stars relative to the
number of subgiant stars, while models with too few binaries with $q
\sim 1.0$ do not extend as far above the old MSTO as observed.  An
example of a model with binaries is shown in Figure~2d.

These preliminary models certainly do not provide a perfect match to
the data; for example, we are unable to reproduce the relatively
narrow color distribution of the observed spur.  Alternative mass
ratio and period distributions, the inclusion of mass transfer
effects, or the effects of binary coalescence could all significantly
modify the model results.  We plan to explore these effects in our
detailed analysis of these data.  Nonetheless, we conclude that the
interpretation of the spur as primarily a binary sequence represents
the only viable interpretation of this feature.

\section{Discussion}

Scl has long been known to have an anomolously red HB for its
predominantly old age and low metallicity (Zinn 1980; Da~Costa 1984;
Mateo 1998).  Our observations confirm that this `second parameter
effect' is present, but only in the inner region of the galaxy.
Oddly, we cannot easily attribute the HB morphology gradient to the
usual suspects; our main-sequence photometry reveals no significant
age or metallicity gradients within Scl.  If confirmed by
spectroscopic abundance determinations, this conclusion casts doubt on
the reliability of the HB as an age indicator of old and
intermediate-age populations (e.g. Da Costa et al. 1996; Mateo 1998).
In Scl, we can speculate that the HB morphology may be related in some
way to the binary component of Scl, the only other population to show
{\it any} radial gradient.  Alternatively, a metallicity gradient may
be present but restricted to a limited number of elements that do
not strongly affect broad-band colors of evolved stars (Majewski et
al. 1999).  A final possibility is that Scl does exhibit age and
metallicity variations, but these largely cancel out in the broad-band
colors for all evolutionary stages. We shall investigate this
(unlikely) possibility in our detailed study of the SFH of Scl.

Binaries have previously been detected both spectroscopically (Queloz
\& Dubath 1995, Armandroff \& Da~Costa 1986) and kinematically
(Olszewski et al. 1996, Armandroff et al. 1995, Hargreaves et
al. 1996) in dSph systems.  Recently, using HST photometry, Gallart et
al. (1999) found that a significant binary population was needed to
explain the detailed morphology of the CMD of Leo~I.  The complex and
temporally extended SFH of that galaxy made it impossible to
disentangle the effects of age and binaries unambiguously. Likewise,
in other dSph with strong intermediate-age components such as Carina
or Fornax, these binaries would be difficult to isolate and would
complicate the interpretation of the CMD.  With its simpler SFH, Scl
provides a cleaner environment to study its binary population
photometrically.

The relatively high inferred binary fraction does not conflict with
the low observed spectroscopic binary fraction of 10-20\%\ (Queloz et
al. 1995; Hut et al. 1992).  It also probably does not significantly
alter conclusions regarding galactic masses based on the kinematics
derived from single-epoch radial velocity measurements of red giants
(Olszewski et al. 1996; Hargreaves et al. 1996) unless the mass-ratio
distribution is very strongly skewed towards unity.  This sort of
distribution seems unlikely in Scl.  Although we infer a significant
number of $q \sim 1.0$ binaries, models containing exclusively
equal-mass binaries do not reproduce our photometric observations of
Scl well.

The central over-concentration of spur stars/binaries towards the
center of Scl may be an important clue in understanding the formation
and evolution of this galaxy. Since mass segregation is implausible
given the relaxation time for this low-density system, the segregation
of binary/RHB stars could be the relic of a time of violent
relaxation, of the accretion of independent systems, or of some
formation process that extended well beyond the present-day boundaries
of the luminous component of the galaxy.  Consequently, systems with
complex substructure such as Fornax (Stetson et al. 1998; see also
Eskridge 1988a,b) and now possibly Scl may contain distinct dynamical
components.  Along with deep, wide-field photometry and spectroscopic
abundances, large-scale spectroscopic studies of the kinematics of
these galaxies will help disentangle the detailed histories of these
deceptively complicated galaxies.

\vspace{1.0em}

DHK and MM thank Gian-Paolo Bertelli and Cesare Chiosi for the use of
the ZVAR synthesis code, and the referee Gary Da Costa for a helpful
review.  MM gratefully acknowledges support from the NSF for this
research.  EKG acknowledges support by NASA through grant
HF-01108.01-98A from the Space Telescope Science Institute, which is
operated by the Association of Universities for Research in Astronomy,
Inc., under NASA contract NAS5-26555.

\newpage

\begin{table}
\caption{\hspace{0.1in} Star Counts and Count Ratios}
\begin{tabular}{lcccccc}
\tableline
\tableline
\multicolumn{7}{c}{Counts} \\
\tableline
 & BHB & RHB & BS & Spur & SGB & Old\\
\tableline
inner & 309 & 275 & 297 & 549 & 1038 & 2671 \\
outer & 62 & 9 & 33 & 36 & 174 & 369 \\
\vspace{0.25pt}\\
\tableline
\tableline
\multicolumn{6}{c}{Count Ratios} \\
\tableline
 & RHB/SGB & BHB/SGB & BS/SGB & Spur/SGB & RHB/Spur \\
\tableline 
inner     & 0.26$\pm0.02$ & 0.30$\pm0.02$ & 0.29$\pm0.02$ & 0.53$\pm0.03$ & 0.50$\pm0.04$ \\
outer    & 0.05$\pm0.02$ & 0.36$\pm0.05$ & 0.19$\pm0.04$ & 0.21$\pm0.04$ & 0.25$\pm0.09$ \\
$\Delta\tablenotemark{a}$ & $7.4\sigma$   & $1.1\sigma$   & $2.2\sigma$   & $6.4\sigma$   & $2.5\sigma$ \\
$T2\tablenotemark{b}$ & $6.4$   & $1.8$   & $2.9$   & $8.8$   & $3.6$ \\
\tableline
\end{tabular}
\tablenotetext{a}{$\Delta = \frac{|R_{inner} - R_{outer}|}{\sqrt{\sigma_{R_{inner}}^2+\sigma_{R_{outer}}^2}}$, where $R$ is the star-count ratio in consideration.}
\tablenotetext{b}{see Da~Costa et al. 1996 for a definition of the T2 statistic}
\end{table}

\newpage

\figcaption[Hurley-Keller.fig1.ps]{Color-magnitude diagrams of the
inner and outer regions of Scl with boxes for reference(see text).
Panel a): CMD of the inner region of Scl with more than 70,000 stars.
Panel b): CMD of the outer region of Scl containing about 30,000
stars.}

\figcaption[Hurley-Keller.fig2.ps]{Possible scenarios to explain the
``spur''. Panel a): the inner CMD with boxes (see text).  Panel b):
the inner CMD with isochrones.  The solid lines show good fits to the
old MSTO (14 Gyr), the spur (8 Gyr), and the young/BS stars ($\ge$4
Gyr) for $z$=0.0004 and a distance modulus of 19.5.  The long-short
dash line is an isochrone with $z$=0.0006 and age=8 Gyr.  The dashed
line is an isochrone with $z$=0.004 and age=4 Gyr.  Panel c): the
inner CMD with a 15 Gyr, $z$=0.0004 isochrone at a distance modulus of
18.8.  Panel d): A simulated CMD with an old population of 15 Gyr with
30\% of the apparent single stars as binaries with a mass fraction
$q\ge$0.7, and a young population generated from a constant SFR from
15 Gyr to 4 Gyr old.  Both populations have $z$=0.0004.}

\newpage
\plotone{Hurley-Keller.fig1.ps}

\newpage
\plotone{Hurley-Keller.fig2.ps}

\end{document}